\documentclass[aps,12pt,final,notitlepage,oneside,onecolumn,
nobibnotes,nofootinbib,superscriptaddress,noshowpacs]{revtex4}
\usepackage{amssymb,amsmath}
\usepackage{mathtext}
\usepackage[cp1251]{inputenc}
\usepackage[T2A]{fontenc}
\usepackage[ukrainian,russian,english]{babel}
\usepackage{bm}
\usepackage{array}
\usepackage{graphicx}
\usepackage{calc}
\usepackage{ifthen}
\usepackage{epsfig}

\begin{document}
\setlength{\baselineskip}{0.7\baselineskip}
\title{Statistical substantiation of introduction of the distributions containing
lifetime as thermodynamic parameter}
\author{V.V. Ryazanov}

\affiliation{Institute for nuclear research of National Academy of
science of Ukraine, Kiev, pr.Nauki, 47, Ukraine \quad
e-mail:vryazan@kinr.kiev.ua}

\begin{abstract}

By means of an inequality of the information and parametrization
of family of distributions of the probabilities, supposing an
effective estimation, introduction of the distributions containing
time of the first achievement of a level as internal thermodynamic
parameter ground.
\end{abstract}


\maketitle

\bigskip
\setlength{\baselineskip}{0.7\baselineskip}


\section{Introduction}

Results of works \cite {Chentsov, Moroz} are applied to a
substantiation of introduction of the distributions containing
time of the first achievement of a level \cite {Ryazanov04,
Ryazanov05, Ryazanov06, Fel, Red}, lifetime in terminology \cite
{Strat61}. The open system is considered, dynamic values under
influence of interaction with an environment become random, the
macroscopical description is possible only. When considering the
finite size systems the finiteness of lifetime seems to be
essential. Influences to which the system is exposed at
interaction with an environment, cause deviations from a normal
steady state and dissipative effects, change a degree of removal
from a stationary state, entropy of system and its lifetime.
Normally functioning system appears to be in a steady
nonequilibrium state characterized by a given deviation from
equilibrium and the entropy production. Each state of a system has
its own lifetime related to fluxes magnitudes and sources
strengths and therefore its deviation from equilibrium.

      Raising lifetime to the realm of the macroscopic
observables in \cite{Ryazanov04,Ryazanov05,Ryazanov06} means more
than the extension of the set of macroscopical, reflects possible
complication of the structure of the phase space that can contain
domains of different behaviour (for example, attractors with
anomalously big lifetimes etc). From the random processes theory
it follows that the existence and finiteness of a lifetime is
provided by the existence of stationary states, physically meaning
the existence of stationary structures. In the theory of random
processes by defining concept of which it is possible to construct
the theory of random processes, the concept of the moments of a
stop to which a lifetime concerns also serves. The events bearing
the information on system, - from the outside or outside, - create
a stream of time in system. The moments of a birth and death are
allocated.

A number of the important phenomena (in the theory of phase
transitions, in chemical reactions, in research of dynamics of
complex biomolecules, calculation of factor of superficial
diffusion in semiconductors, in researches of nucleus, elementary
particles, in spin glasses, spectroscopy, in systems with traps,
in the theory of metastable states, etc.) it is physically caused
by finiteness of a lifetime of system, opportunities of those
greater fluctuations in behaviour of the statistical system,
having small probability, which, as a rule, are not considered at
the description of evolution of system.

The lifetime is the time period till the moment of the first
(random) achievement \cite {Fel, Red} a certain level (for
example, zero level) by a random process  $y (t)$ for
macroscopical parameter of system (for example, its energy or
number of particles). Lifetime is the subordinated random process
in terminology of the theory of random processes \cite {Fel}; the
behaviour of lifetime $\Gamma$ is defined by behaviour of random
process $y (t)$:

\begin{equation}
\Gamma_{x}=inf\{t:y(t)=0\}, \quad y(0)=x>0.
 \label{time}
\end{equation}

The lifetime depends on the energy of a system, its size, fluxes
of energy. Therefore a system exchanges with thermostat the
energy, the particle number, but not the "lifetime". The lifetime
is a macroscopical parameter characterizing the system and its
interaction with the environment. The lifetime is an observable,
well-defined and physically well understandable quantity
reflecting important system peculiarities. As a lifetime waiting
times essential at the description of processes of fractional
diffusion can be considered.

We consider a lifetime, an interval of time, defined evolution of
the system, depending from its properties. And back: properties of
system depend on its lifetime. The important characteristic of any
system is its age. In thermodynamics and the statistical physics
it is supposed, that statistical systems tends to equilibrium
state. We consider the open systems for which the equilibrium
state is badly certain. But all real physical systems of the
finite sizes have final lifetime, that essentially influences
their properties and properties of their environment. It is the
universal physical characteristic. We do not consider ideal system
of the infinite sizes, we do not do thermodynamic limiting
transition. Many systems degenerate, not reaching equilibrium.
Therefore in more representative physical value, than time of
achievement of equilibrium, is lifetime.

In works of the Brussels school \cite {Prig, Prig3} the
constructive role of irreversibility, that nonequilibrium can be a
source of the order is emphasized. In \cite {Prig} by
consideration of a problem being and becoming the operator of
internal time is entered. It is possible to draw an analogy
between it and lifetime. Initial conditions thus arise as an
outcome of previous evolution of the system. Both lifetime and
internal time from work \cite {Prig} serve for the description of
properties dissipative systems with age inherent in them and
allocation of a direction of current of time. Lifetime is certain
by trajectory with the allocated point of the moment of
achievement of a zero level in the future. Prigogine's approach
\cite {Prig} and the concept of lifetime unites presence of some
allocated moment, event, as integral part of the probabilistic
description and occurrence new among already present systems,
elements, interactions. In the present work it is the moment of
destruction of the system, and in \cite {Prig} it is bifurcation,
transition to chaos, as consequence of instability, i.e. other
pole stationary-relaxation evolution of the system described, for
example, by model of storage \cite {Pr, Ryaz93, Sp}. Let's note,
that the description offered here supposes natural generalization
on cases of loss of stability, transition to state among which is
not present stationary \cite {Ryaz93}. Times before transition in
these states which are precisely set for model of storage are
defined also.

In the present work strict results of mathematical statistics (for
example, \cite {Chentsov} and \cite {Moroz}), the families of
distributions of probabilities connected with dependence on some
parameter are used. We consider a situation when as this parameter
lifetime of system acts. As examples of such distributions serve
quasi-equilibrium distributions \cite {Zub80,Zub99} and the
nonequilibrium statistical operator (NSO) \cite {Zub71, Zub80,
Zub99}. In work \cite {Ryazanov01} logarithm NSO $\rho (t)$ \cite
{Zub71, Zub80, Zub99} is interpreted as averaging of the logarithm
of quasi-equilibrium distribution $\rho _ {q}$ \cite {Zub80,
Zub99} from various time arguments on distribution $p_{q}(u)$ of
lifetime of system (time of the first achievement of a level):

\begin{equation}
\ln \rho(t)=\int_{0}^{\infty}p_{q}(u)\ln \rho_{q}(t-u,-u)du
 ,\, \label{nso}
\end{equation}
where $u=\Gamma=t-t_{0}$ is a random variable of lifetime of the
system, $t$ is a present moment of time, $t _ {0}$ is a random
variable of the initial moment of time, "birth" of the system. The
value $u=t-t_{0}$ is equal also to the random moment of the first
achievement of a zero level \cite {Fel, Red} during the moment
$t_{0}$ in return time, at $t\mapsto-t$, Figure 1, (\ref {time}).
\bigskip
\begin{figure}
\begin{center}
\epsfig{figure=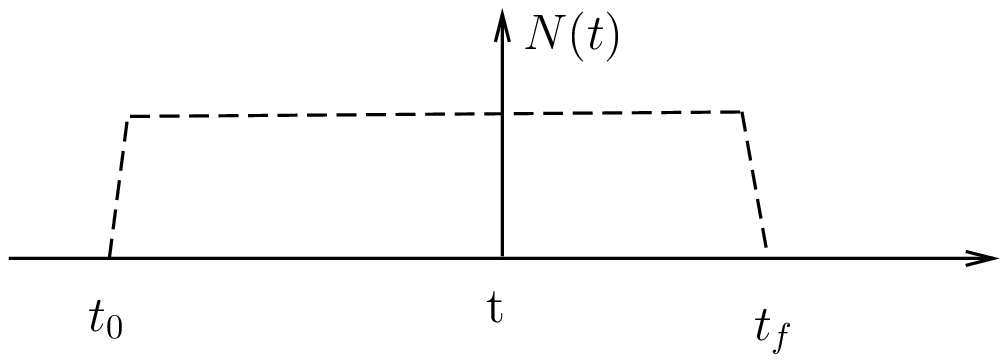, width=1\textwidth, clip}
\end{center}
\caption{ }
\label{fig1}
\end{figure}
If $p_{q}(u)=\varepsilon exp\{-\varepsilon u\}$, distribution
$p_{q}(u)$ has an exponent appearance with
$\varepsilon=1/\langle\Gamma\rangle$, where
$\langle\Gamma\rangle=\langle t-t_{0} \rangle$ is average lifetime
of the system, from (\ref {nso}) is received NSO in the form of
Zubarev \cite {Zub71, Zub80, Zub99}.

\section{The distributions containing lifetime as thermodynamic parameter}

The smooth families depending on parameter $\overrightarrow {x}$,
are considered in \cite {Chentsov} and in \cite {Moroz} together
with an inequality of the information. In \cite {Chentsov} the
Theorem 15.3 is proved.

Let for family of distributions of probabilities
$P\{\cdot|\overrightarrow {x}\},\overrightarrow{x}=(x_{1}...,
x_{n})$, set some local display $n$-dimensional space
$\overrightarrow {X}$, exists vector statistics $\overrightarrow
{\tau} (\omega)$, setting nonbiased estimation of parameter of
family

\begin{equation}
M_{\overrightarrow{x}}\tau(\omega)=\langle
P(\overrightarrow{x}),\overrightarrow{\tau}\rangle
=\int\tau(\omega)P\{\overrightarrow{x},d\omega\}=\overrightarrow{x}.
\label{parsem}
\end{equation}
If the family is differentiated at $\overrightarrow {x} =
\overrightarrow {\theta}$ with the second moment, $\overrightarrow
{W} = \overrightarrow {W} (\overrightarrow {\theta})$ is its
Fisher's information matrix (\ref {M1})

\begin{equation}
\frac{\partial \ln p(\omega ; \overrightarrow{x})}{\partial x_j}=
r^{j}(\omega; \overrightarrow{x}), \qquad \frac{\partial^{2}{\ln
p(\omega; \overrightarrow{x})}}{\partial x_j
\partial x_k}= r^{jk}(\omega;
\overrightarrow{x})\, ,
 \label{r}
\end{equation}
\begin{equation}
-M_{\overrightarrow{x}}r^{jk}(\omega;\overrightarrow{x})=M_{\overrightarrow{x}}r^{j}(\omega;
\overrightarrow{x})r^{k}(\omega;
\overrightarrow{x})=\overrightarrow{\omega}^{jk}(\overrightarrow{x}),\,
 \label{M1}
\end{equation}
$\overrightarrow{w^{ij}}(\overrightarrow{\theta})=
(P'_{i}(\overrightarrow{\theta}),P'_{j}(\overrightarrow{\theta}))_{P(\theta)},
\overrightarrow{W}(\overrightarrow{\theta})=(w^{ij})_{i,j=1}^{n}$,
$\overrightarrow{W}^{-1}=\overrightarrow{V}=(v_{ij})_{i,j=1}^{n}$,
$p(\omega; \overrightarrow{x})$ is density of probability of
family $P\{\cdot|\overrightarrow{x}\}$, $\ln p(\omega;
\overrightarrow{x})$ is function of likelihood, then

\begin{equation}
\sum_{i,j=1}^{n}y^{i}y^{j}M_{\theta}[\tau_{i}(\omega)-\theta_{i}]
[\tau_{j}(\omega)-\theta_{j}]\geq
\sum_{i,j=1}^{n}y^{i}y^{j}\overrightarrow{v}_{ij} , \label{neinf}
\end{equation}
where $\overrightarrow{y}=(y^{1},...,y^{n})$ is any vector of the
conjugate space $Y=X^{X}$.

The inequality (\ref {neinf}) turns to the equality equivalent
$n^{2}$ correlations

\begin{equation}
\overrightarrow{M}_{\theta}[\tau_{i}(\omega)-\theta_{i}]
[\tau_{j}(\omega)-\theta_{j}]=\overrightarrow{v}_{ij} ,
\label{ekv}
\end{equation}

in only case when, when $P_{\theta}$-almost everywhere

\begin{equation}
\sum_{i=1}^{n}\overrightarrow{v}_{ij}
\frac{dP'_{i}(\overrightarrow{\theta})}{dP(\overrightarrow{\theta})}(\omega)=
[\tau_{j}(\omega)-\theta_{j}], \quad  j=1,...,n , \label{uslov}
\end{equation}
or when $P _ {\theta} $-almost everywhere

\begin{equation}
r^{i}(\omega;\overrightarrow{\theta})=
\frac{dP'_{i}(\overrightarrow{\theta})}{dP(\overrightarrow{\theta})}(\omega)=
\sum_{j=1}^{n}\overrightarrow{w}_{ij}[\tau_{j}(\omega)-\theta_{j}],
\quad  j=1,...,n . \label{uslov1}
\end{equation}

Let the family of distributions of probabilities
$P\{\cdot|\overrightarrow {x}\}$ of parameter $\overrightarrow
{x}$ is those, that the inequality of the information (\ref
{neinf}) for it is meaningful and fair. The statistics $\tau
(\omega)$ refers to as an effective estimation of parameter
$\overrightarrow {x}$, if it nonbiased and an inequality of the
information (\ref {neinf}) for it addresses in equality at each
value of parameter. Conditions of efficiency look like (\ref
{uslov}) and (\ref {uslov1}). The necessary and sufficient
condition of transformation of an inequality (\ref {neinf}) in
equality is transformed to a condition for density of a kind \cite
{Moroz}

$$
\exists \Omega': P_{\theta}(\Omega')=1; \quad \forall\omega\in
\Omega', \quad
\tau(\omega)-\theta=w^{-1}(\theta)\frac{dP'_{\theta}}{dP_{\theta}}(\omega).
$$

In \cite {Chentsov} the Theorem 15.4 is proved. If parametrization
$\overrightarrow {x} = (x _ {1}..., x _ {n})$ of family of
distributions of probabilities unitary differentiated with the
second moment $P\{\cdot|\overrightarrow{x}\}$ supposes an
effective estimation $\overrightarrow {\tau} (\omega)$ the family
is geodetic (exponent) with directing statistics $\overrightarrow
{\tau} (\omega) = (\tau _ {1} (\omega)..., \tau _ {n} (\omega))$,
i.e. it is set by density:

\begin{equation}
\frac{dP_{\overrightarrow{x}}}{dP_{0}}(\omega)=p(\omega;\overrightarrow{x})=
exp[\sum_{j}s^{j}(\overrightarrow{x})\tau_{j}(\omega)-\Psi((\overrightarrow{s}(\overrightarrow{x}))]
,\, \label{exp2}
\end{equation}
where $exp[\Psi((\overrightarrow{s}(\overrightarrow{x}))]$ is a
normalizing divider, and $\overrightarrow {x}$ is its natural
parametrization mutually-is unequivocally adequating to canonical
parametrization $\overrightarrow {s} = (s ^ {1}..., s ^ {n})$,
$\overrightarrow {x} = \overrightarrow {M} _ {\overrightarrow {s}}
\overrightarrow {\tau} (\omega)$.

Parametrization $\overrightarrow {x}(P_{s})=(x_{j})_{j=1}^{n}$ of
exponent (geodetic) family (\ref {exp2}) refers to natural,
adequating to the given canonical representation (\ref {exp2}), if

$$
x_{j}=X_{j}(\overrightarrow{s})=\overrightarrow{M}_{\overrightarrow{s}}\tau_{j}(\omega)=
\int_{\Omega}\tau_{j}(\omega)p(\omega,\overrightarrow{s})\mu\{d\omega\}.
$$

If to consider (\ref {exp2}) as distribution of statistical
physics, the value of parameter of distribution appears equal to
average value of random internal thermodynamic parameter \cite
{Strat82} (observable value, as $\langle P _ {n} \rangle$ in (\ref
{relev})). Thus, the value $\tau (\omega)$ is internal
thermodynamic parameter with average value $x$ parameter of
distribution. The parameter of distribution acts in a role of
observable value. In that specific case theorem 15.4 when the
effective estimation of parameter corresponds to lifetime of the
system $\Gamma=t-t _ {0}$ (\ref {time}), for example, for family
of quasi-equilibrium distributions (\ref {relev}), \cite {Zub99},
depending from $\Gamma$,

\begin{equation}
\ln
\rho_{q}(t_{1},t_{2})=-\Phi(t_{1})-\sum_{n}F_{n}(t_{1})P_{n}(t_{2})
 ,\, \label{relev}
\end{equation}
where, as in (\ref {nso}), $t_{1}=t-u,  t_{2}=-u;
u=\Gamma=t-t_{0}$, fair will be representation joint for energy
$E$ and lifetime $\Gamma$ the distribution, written down in \cite
{Ryazanov04, Ryazanov05, Ryazanov06} in the form

\begin{eqnarray}
P(E,\Gamma)=\omega(E,\Gamma)\rho(z;E,\Gamma); \quad
\rho(z;E,\Gamma)=exp\{-\beta E-\gamma\Gamma\}/Z(\beta,\gamma);\nonumber \\
Z(\beta,\gamma)=\int e^{-\beta E-\gamma\Gamma}dz= \int \int
e^{-\beta E-\gamma\Gamma}\omega(E,\Gamma)dEd\Gamma,
 \label{life}
 \end{eqnarray}
where $\Gamma=t-t _ {0}$ is random lifetime (\ref {time}), time of
the first achievement of a zero level \cite {Fel, Red} in return
time, $\gamma$ is the intensive thermodynamic parameter conjugate
by thus thermodynamic variable, and the generalized structural
factor $\omega (E, \Gamma)$ satisfies to a condition $\int \omega
(E, \Gamma) d \Gamma = \omega (E)$ (the usual structural factor
\cite {Klim}). In work \cite {Ryaz05} by means of distribution
(\ref {life}) expression for distribution on energies of neutrons
in a nuclear reactor corresponding experimental results is
received. Some other possible applications of this distribution
are described in \cite {Ryazanov05, Ryazanov06}.

Expressions (\ref {relev}) for $\rho _ {q}$ correspond to
exponential family (\ref {exp2}) and coincides with it at

\begin{equation}
\rho_{q}(t;\omega)=p(\omega;\overrightarrow{\theta})/p_{0}(\omega);
\quad \overrightarrow{\theta}=\overrightarrow{x}=u=t-t_{0},
\label{sravn}
\end{equation}
$$
\Phi(t-u)=\Psi(\overrightarrow{s}(\overrightarrow{\theta})), \quad
P_{n}=\tau_{j}(\omega); \quad
F_{j}(t-u)=-s^{j}(\overrightarrow{\theta}) .
$$
For distribution (\ref {life}) as parameter $x$ from theorems
15.4, 23.6 \cite {Chentsov} or theorems 9.3 \cite {Moroz} average
value $\langle \Gamma \rangle $ (unlike a correlation (\ref
{sravn}) where as parameter of distribution not average value $
\Gamma $ serves) serves; $\tau _ {2} (\omega) = \Gamma$, the value
$x _ {2} =M _ {s} \tau _ {2} (\omega)$ from the theorem 15.4 is
equal

$$
\langle\Gamma\rangle=\int \frac{e^{-\beta
E-\gamma\Gamma}\omega(E,\Gamma)}{Z(\beta,\gamma)}d\Gamma=-\frac{\partial
\ln Z(\beta,\gamma)}{\partial \gamma}_{|\beta}.
$$

In relation to value of lifetime $\Gamma=t-t _ {0}$ the term
"thermodynamic parameter"   instead of "a dynamic variable"  (as,
for example, energy), as $\Gamma$ is the statistical value,
directing statistics in terminology \cite {Chentsov}, satisfying
to definition of thermodynamic parameter (for example, \cite
{Leon, Strat82}) is used: any functions $B _ {\alpha} (z)$ from
dynamic variables $z = (q _ {1}..., q _ {N}; p _ {1}..., p _ {N})$
($z = \omega$ in the expressions resulted above), having
macroscopical character, by definition are random internal
thermodynamic parameters. Dependence $\Gamma (z)$ is visible, for
example, from the equations for distribution of lifetime in markov
model (Pontryagin's equations
 \cite {Strat61, Tich}) which are conjugate to the
equations of type the Fokker-Plank. The values of lifetime is
influenced attractors, metastable states, phase transitions, and
other physical features of system depending from $z$. Dependence
$\Gamma$ from $z$ can be set in the form of $\Gamma (z) =t-t _
{0}, z \neq 0; \Gamma (z) =t _ {f}-t _ {0}, z=0, t> t _ {f}$
(Figure 1); $t _ {0} (z) =t, z=0; t _ {0} (z) =t _ {0}, z \neq 0
$. Averaging of value $t _ {0}$ is defined by statistics of a
random variable $\Gamma$. Parameters of distribution (\ref {exp2})
$s$ and $\Psi$ in \cite {Chentsov, Moroz} are expressed through
Fisher's information matrix (\ref {M1}).

Let's receive expressions (\ref {exp2}), (\ref {life}). The
quasi-equilibrium distribution (\ref {relev}) we shall write down
for simplicity with one parameter $P _ {1} =E$. Then
\begin{equation}
\ln \rho_{q}(t-u,-u)=-\beta(t-u)E(-u)-\ln Z(\beta(t-u)) ,
 \label{onepar}
\end{equation}
where $u=t-t _ {0} = \Gamma$ is a random variable of lifetime.
Distribution (\ref {onepar}) depends on parameter $\Gamma$. We
shall assume, that it depends and from $\langle \Gamma \rangle$
(through $\beta (\langle \Gamma \rangle)$, for example, at $\Gamma
= \langle \Gamma \rangle, t - \Gamma = \langle \Gamma \rangle$).
Besides the value $\beta$ in (\ref {onepar}) depends from $\langle
E \rangle$. Therefore the parameter $\overrightarrow {x}$ in the
theorem 15.4 is equal
\begin{equation}
\overrightarrow{x}=(\langle E\rangle,\langle\Gamma\rangle); \quad
\overrightarrow{\tau}=(E,\Gamma) .
 \label{2par}
\end{equation}

Let's notice, that any other dependence of function of
distribution from $\Gamma$ and $\langle \Gamma \rangle$ also
satisfies to conditions of the theorem 15.4. Let's write down for
distribution (\ref {onepar}) the values entering into a condition
of an effective estimation (\ref {uslov1}) under condition of
$\frac {\partial E (-\Gamma)} {\partial \langle \Gamma \rangle}
=0$ which corresponds to a stationary case, and also considered in
\cite {Chentsov} situations, when in (\ref {exp2}) $s^{j}
(\overrightarrow {x})$, and $\tau _ {j} (\omega)$:
\begin{eqnarray}
\frac{\partial \ln \rho_{q}}{\partial \langle
E\rangle}_{|\langle\Gamma\rangle}=\frac{\partial \beta}{\partial
\langle E\rangle}_{|\langle\Gamma\rangle}[\langle
E\rangle-E]; \nonumber \\
\frac{\partial \ln \rho_{q}}{\partial
\langle\Gamma\rangle}_{|\langle E\rangle}=-\frac{\partial
\beta}{\partial \langle\Gamma\rangle}_{|\langle
E\rangle}E-\frac{\partial \ln
Z(\beta(t-\langle\Gamma\rangle)}{\partial \beta}\frac{\partial
\beta}{\partial \langle\Gamma\rangle}_{|\langle
E\rangle}=\frac{\partial \beta}{\partial
\langle\Gamma\rangle}_{|\langle E\rangle} [\langle E\rangle-E]
\nonumber .
\end{eqnarray}

According to (\ref {life}) we shall enter thermodynamic parameter
$\gamma$ and we shall consider dependences $\langle E (\beta,
\gamma) \rangle, \langle \Gamma (\beta, \gamma) \rangle$. Then
elements of an information matrix of Fisher (\ref {M1}) are equal
$$
w^{11}=(\frac{\partial \beta}{\partial \langle
E\rangle}_{|\langle\Gamma\rangle})^{2}(\langle
E^{2}\rangle-\langle E\rangle^{2})=
\frac{D_{EE}}{D^{2}}(D_{\Gamma\Gamma})^{2}; \quad \frac{\partial
\beta}{\partial \langle
E\rangle}_{|\langle\Gamma\rangle}=-D_{\Gamma\Gamma}/D;
$$
$$
w^{12}=w^{21}=-\frac{D_{EE}}{D^{2}}({D_{\Gamma\Gamma}})({D_{\Gamma
E}}); \quad w^{22}=\frac{D_{EE}}{D^{2}}(D_{\Gamma E})^{2}; \quad
\frac{\partial \beta}{\partial \langle\Gamma\rangle}_{|\langle
E\rangle}=D_{E\Gamma}/D;
$$
$$
D_{EE}=-\frac{\partial \langle E\rangle}{\partial
\beta}_{|\gamma}=\langle E^{2}\rangle-\langle E\rangle^{2}; \quad
D_{\Gamma E}=D_{E\Gamma}=-\frac{\partial
\langle\Gamma\rangle}{\partial \beta}_{|\gamma}=-\frac{\partial
\langle E\rangle}{\partial \gamma}_{|\beta}=\langle\Gamma
E\rangle-\langle\Gamma\rangle\langle E\rangle;
$$
$$
D_{\Gamma\Gamma}=-\frac{\partial \langle \Gamma\rangle}{\partial
\gamma}_{|\beta}=\langle\Gamma^{2}
\rangle-\langle\Gamma\rangle^{2}; \quad
D=D_{\Gamma\Gamma}D_{EE}-D_{\Gamma E}^{2} .
$$

The condition (\ref {uslov1}) becomes
\begin{equation}
\langle E\rangle-E=L(\langle \Gamma\rangle-\Gamma); \quad
L=\frac{D_{EE}}{D_{\Gamma E}} .
 \label{con2}
\end{equation}

Thus, the condition of that that quasi-equilibrium distribution
(\ref {onepar}) becomes (\ref {exp2}) (or (\ref {life}) at
parameter of a kind (\ref {2par})) looks like (\ref {con2}).

Let's show, how can to be connected distributions (\ref {onepar})
and (\ref {life}). Let in (\ref {onepar}) $\beta (t-u) = \beta
(\langle E \rangle, \langle \Gamma \rangle) = \beta _ {1} + \beta
_ {2}$. Using (\ref {con2}), we shall write down $\beta _ {2} E =
\beta _ {2} [-L (\langle \Gamma \rangle - \Gamma) + \langle E
\rangle]$. Then at $\beta _ {1} = \beta, \beta _ {2} L = \gamma,
\ln Z (\beta, \gamma) = \ln Z + \beta _ {2} [\langle E \rangle -
\langle \Gamma \rangle L]$, from distribution (\ref {onepar}) we
receive distribution (\ref {life}). Thus distributions (\ref
{onepar}) and (\ref {life}) differ different values $\beta$. In
many stationary cases $D _ {\Gamma E} =D _ {E \Gamma} = \langle
\Gamma E \rangle - \langle \Gamma \rangle \langle E \rangle \sim
0$, $\gamma \sim 1 / \langle \Gamma \rangle$. Therefore the value
$L$ accepts great values, $\beta _ {2} <<\beta _ {1}, \beta
(\langle E \rangle, \langle \Gamma \rangle) \simeq \beta _ {1}$,
and the specified distinction of values $\beta$ in these cases can
be neglected.

If in the correlation (\ref {nso}) $p_{q}(u)=\varepsilon
exp\{-\varepsilon u\}$, and $\hat {\sigma} (t-u,-u) \simeq \hat
{\sigma} (t)$, (\ref {nso}) becomes

\begin{equation}
\ln \rho(t)\simeq \ln
\rho_{q}(t,0)+\hat{\sigma}(t)\langle\Gamma\rangle ,
 \label{dew}
\end{equation}
as $\varepsilon^{-1} = \langle \Gamma \rangle = \langle t-t_{0}
\rangle$. In this case $\ln \rho _ {q} = \int \hat {\sigma} (t) du
= \hat {\sigma} (t) u+C$, where $C$ does not depend from $u$, and
we receive for $\rho _ {q}$ distribution of exponent a kind on $u$
(though it only a special case received above results).
Substitution of this expression for $\ln \rho _ {q}$ in (\ref
{nso1}) \cite {Ryazanov01}

\begin{equation}
\ln \rho(t)=\int_{0}^{\infty}p_{q}(u)\ln \rho_{q}(t-u,-u)du=\ln
\rho_{q}(t,0)-\int_{0}^{\infty}(\int p_{q}(u)du)
\hat{\sigma}(t-u,-u)du, \label{nso1}
\end{equation}
$\hat {\sigma} (t-u,-u) = \partial \ln \rho _ {q} (t-u,-u) /
\partial u$,
leads to a correlation (\ref {dew}) at $C = \ln \rho _ {q} (t, 0)
$. Distribution (\ref {dew}) has been received in \cite {Dew} by
other methods and applied to geophysical problems. Then
expressions for NSO (\ref {nso}), (\ref {nso1}) correspond to
expression (\ref {life}), where
$\ln\rho(t)=\ln\rho_{q}(t,0)+\ln[Z(\beta)/Z(\beta,\gamma)]-\gamma\Gamma$,
where $\rho_{q}(t,0)=\exp\{-\beta E\}/Z(\beta),
Z(\beta)=\int\exp\{-\beta E\}dz$; the elementary case of a
correlation (\ref {relev}) is considered, at $n=1$. But in (\ref
{life}) instead of multiplier $\langle \Gamma \rangle \hat
{\sigma}$ containing average value $\langle \Gamma \rangle$,
appears a random variable of lifetime $\Gamma$ and Lagrange
multiplier $\gamma$; the value $\langle \Gamma \rangle \hat
{\sigma}$ is replaced with fuller expression $ [Z (\beta)/Z
(\beta, \gamma)] - \gamma \Gamma $. In (\ref {life}) averaging on
distribution of lifetime (as to parameter of distribution) in NSO
(\ref {nso}), (\ref {nso1}) is replaced with use of random
thermodynamic parameter of lifetime $\Gamma $.

The parameter $\gamma$, connected with value $\hat {\sigma}$,
depends on a specific task. So, for prompt neutrons in a nuclear
reactor \cite {Ryaz05} $\gamma=-k/\Gamma_{0}$, where $k$ is
multiplicative factor of neutrons, $\Gamma _ {0}$ is average
lifetime of neutrons without perturbation, in not making multiple
copies environment. In problems of phase synchronization \cite
{Strat61, Tich} $\gamma=\frac{\Delta}{8\pi^{2}D}\frac{[1+exp\{2\pi
D_{0}\}]} {\exp\{\pi D_{0}\}|I_{iD_{0}}(D)|^{2}}$, where $\Delta$
is a strip of synchronization (keeping) of the connected
generators, $\Delta _ {0}$ is size of initial discomposure of
generators $D _ {0}/D = \Delta _ {0} / \Delta$, $I _ {iD _ {0}}
(D)$ is tabulated Bessel function with imaginary argument and an
imaginary index.

Above in this section as parameter of distribution the value
$\langle \Gamma \rangle$ was considered, though as parameter, on
which obviously depends quasi-equilibrium distribution (\ref
{relev}), the value of age of system $\Gamma=t-t _ {0}$ was
considered. Here too it is possible to consider value $\Gamma=t-t
_ {0}$ as parameter of distribution, having assumed, that this
value represents average value of a random variable $\tau$, i.e.
$\Gamma = \langle \tau \rangle$. To prove introduction of value
$\tau$ it is possible, for example, following work \cite {Prig},
where the operator of time (fair and for a classical case),
representing fluctuating time is entered; "usual"    time enters
the name in \cite {Prig} as average value of the operator of time.
One more opportunity of a ground of introduction of a random
variable $\tau$ is connected with the initial moment $t _ {0}$. If
value $t _ {0}$ is already average on some parameters, $t _ {0} =
\langle \tau _ {0} \rangle$, and $\Gamma = \langle \tau=t - \tau _
{0} \rangle$. Thus, in expressions (\ref {life}), (\ref {con2}),
$w {ij}$, instead of value $\langle \Gamma \rangle$ as average
value appears value $\Gamma$, and the random variable $\Gamma$ is
replaced with a random variable $\tau$. As in a stationary case,
at $\partial \langle E \rangle / \partial \Gamma=0$ and \quad
$\partial /
\partial t = \partial / \partial \Gamma$, \quad $\frac {\partial
\beta} {\partial \Gamma} = \frac {\partial \beta} {\partial
\Gamma} _ {| \langle E \rangle}$, $\partial \ln \rho _ {q} /
\partial \Gamma _ {| \langle E \rangle st} = \sigma _ {st}$. Thus
\quad $\langle \sigma E\rangle_{st}=-D_{E\Gamma}D_{EE}/D_{|st},
\quad
\langle\sigma^{2}\rangle_{|st}=D^{2}_{E\Gamma}D_{EE}/D^{2}_{|st},
\quad \langle\sigma\tau\rangle_{st}=-D_{E\Gamma}^{2}/D_{|st},
\quad D_{EE |st}=\langle\sigma
E\rangle^{2}/\langle\sigma^{2}\rangle_{|st}$ (this correlation
gives expression for a thermal capacity ), $\quad D_{\Gamma\Gamma
|st}=\langle\sigma\tau\rangle(\langle\sigma\tau\rangle-1)/\langle\sigma^{2}\rangle_{|st},
\quad D_{E\Gamma |st}=\langle\sigma
E\rangle\langle\sigma\tau\rangle/\langle\sigma^{2}\rangle_{|st},
\quad \frac{\partial \beta}{\partial \Gamma}_{|\gamma
st}=\frac{\partial \beta}{\partial t}_{|\gamma st}=-1/D_{E\Gamma
|st}=-\langle\sigma^{2}\rangle/\langle\sigma\tau\rangle\langle\sigma
E\rangle_{|st}; \quad L=\langle\sigma
E\rangle/\langle\sigma\tau\rangle_{|st}; \quad \partial \langle
E\rangle/\partial t_{|\gamma st}=D_{EE}/D_{E\Gamma|st}=L; \quad
\partial \langle E\rangle/\partial t_{|\beta
st}=D_{E\Gamma}/D_{\Gamma\Gamma |st}=\langle\sigma
E\rangle/(\langle\sigma\tau\rangle-1)_{|st}=L\langle\sigma
\tau\rangle/(\langle\sigma\tau\rangle-1)_{|st}; \quad
\partial \gamma/\partial t_{|\beta st}=(\partial\Gamma/\partial
\gamma_{|\beta
st})^{-1}=(\langle\tau^{2}\rangle-\langle\tau\rangle^{2})^{-1};
\quad \partial \beta/\partial \Gamma_{|\langle E\rangle
st}=\partial \beta/\partial t_{|st}=
-\langle\sigma^{2}\rangle/\langle\sigma E\rangle_{|st}; \quad
\partial \beta/\partial \langle E\rangle_{|\Gamma
st}=\langle\sigma^{2}\rangle(\langle\sigma\tau\rangle-1)/\langle\sigma
E\rangle^{2}_{|st}$.

\section{The conclusion}

For lifetime in a method of the nonequilibrium statistical
operator \cite {Zub71} the past which has been last lifetime
$t-t_{0}$, age of system, Figure 1, where $t_{0}$ is the moment of
a birth of system, $t$ is a present moment of time, $t_ {f}$ is
the moment of destruction of system, $N(t)$ is number of particles
in system during the moment $t$, $t_{f}-t$ is future lifetime, is
considered. Full lifetime, the period of employment in terminology
of the theory of queuing, is equal $t_{f}-t_{0}=t-t_{0}+t_{f}-t$
is the sum of the past and the future. Averaging on the last
lifetime known unlike the uncontrollable future, communicates in
\cite {Zub71} with causality, a sign on a source in the equation
for NSO and signs on local entropy production.

It is shown, that quasi-equilibrium distribution (\ref {relev}),
depending on parameter $\Gamma$, it is possible to write down in
exponent a kind (\ref {exp2}), (\ref {life}). Generally
distribution (\ref {life}) is entered irrespective of
distributions (\ref {relev}) and (\ref {onepar}) and from NSO, and
argument $\Gamma$ in it can be the value which is distinct from
$t-t _ {0}$. Therefore it is difficult to tell, what distribution
more the general: (\ref {life}) or NSO (\ref {nso}).

In \cite {Leon} it is marked, that the nonequilibrium state is
characterized by additional macroscopical parameter in the
description of system. In works \cite {Kirk} it was marked, that
the state of system in a present situation of time depends on all
previous evolution of the nonequilibrium processes developing it
and, accordingly, from time of the last life of system, its age.
In works \cite {Zub71, Zub80, Zub99} dynamics of behaviour of
system and the contribution of correlations during all past of
system is considered. Therefore we choose random lifetime of
system as thermodynamic parameter.

The structure of phase space generally accepts rough (fractal) a
kind caused by explosive spasmodic character of stochastic
dynamics in it. Lifetime is macroscopical value. As the open
systems for which during their interaction with an environment
there is the "exchange"    in dynamic values breaking their
conservation and doing them random, introduction of lifetime as a
basic random variable, it is obviously important also a necessary
element of the correct description of the nonequilibrium phenomena
are considered.

It is mean explaining what is understood under the notion of a
first-passage time or lifetime. Arbitrary systems are formed from
elements, particles, or elementary objects. These elements enter
any system and leave it. It is possible to formulate the notion of
a lifetime with mathematical strictness; one thus understands the
time period at which the elements are present in a system. For
example, if we consider a small volume cell of a gas, its lifetime
is the period till all molecules occasionally leave this cell. Are
examples borrowed from chemical kinetics are times at which the
number of molecules of certain kind becomes zero, time of the
dissociation of a two-atom molecule, the age of a (biological)
system, residence times of grains in sand-pile models etc.
Stratonovich in \cite{Strat61} used the term  "lifetime"    as
terminus techniques with respect to the number of phenomena
considered. Following synonyms are encountered by us: the first
passage time (for some given level), escape time, first-exit time,
busy period (in the queuing theory) etc. The physical value of
lifetime of system, as an interval of time during which at system
there is a nonzero number of elements of which the statistical
system is made is entered into consideration. This value depends
both on internal properties of system, and from external
influences and is generally a random variable. The lifetime is
connected with characteristic time intervals of system (time of
collisions, time of mixture, time of a relaxation, etc.) and with
fluxes (also time characteristics). The theory with lifetime
generalizes approaches \cite {Zub71, Zub80, Zub99},
R.L.Stratonovich \cite {Strat82} and the extended nonequilibrium
thermodynamics \cite {Leb}. The analogy to internal time \cite
{Prig} is spent. Introduction of lifetime opens opportunities for
strict studying the nature of time and allows to receive the
description of transfer processes and other nonequilibrium
phenomena. So, in works \cite {Atman} lifetime of finite mods of
laser systems connection with correlation time of amplitudes of
mods. Average lifetime of laser mods is determined experimentally
on spectra of absorption. The connection of value $\Gamma$ is
formally essential to a developed method, as subordinated process,
with the basic random process $E$. But the concept of lifetime has
as well deep physical sense, uniting Newton the approach by
absolute time and ideas about a matter generating time. The
parameter of lifetime unites the features inherent in usual
dynamic variables like energy, number of particles, and to
coordinate variables of type of time. The mathematical party of
introduction of lifetime consists in reception of the additional
information on stochastic process, except for knowledge of its
stationary distribution, on stationary properties detailed of it
the subordinated process. Irreversibility in this approach appears
as consequence of the assumption of existence and finiteness of
lifetime of system, detailed the moments of a birth and
destruction of system. Physical values, for example, a thermal
capacity of system, can essentially depend on lifetime. So if to
assume, that in system it is realized lifetimes belonging only
some area $\Gamma _ {m}$ (it, for example, it is fair for
metastable states) the contribution to the measured value is given
with values $\Gamma \in \Gamma _ {m}$, those areas of phase space,
where $ \Gamma (E) \gg t _ {obs} $ are accessible only (or, on the
contrary, $\ll t _ {obs}$ are accessible only; instead of time of
supervision $t _ {obs}$ there can be other characteristic scale)
average value, for example, energy is given by expression
$<E>=\int_{0}^{\infty}EdE\int_{\Gamma\in
\Gamma_{m}}d\Gamma\omega(\Gamma,E)\rho(z,E,\Gamma)$. Thus as
$\Gamma _ {m}$, and expression under last integral, can depend on
temperature as on a regular basis, and with features,
characteristic for phase transitions, when jump the whole areas of
phase space become accessible (or inaccessible) system. Dependence
$\Gamma _ {m}$ from temperature is connected with nonequilibrium
phase transitions. Detailed mesoscopical description can be
received with use of obvious stochastic models of system, for
example, diffusion type \cite {Strat82} or stochastic processes of
storage \cite {Pr, Ryaz93, Sp}. The last can be considered as the
models of the system raised by generalized noise. In them
interaction with thermostat (reservoir), and also the moments of
degeneration and loss stationarity is considered at performance of
the certain conditions.

Introduction of lifetime as thermodynamic parameter speaks that
real systems possess finite lifetime that essentially influences
their properties and properties of their environment. Lifetime of
system is represented in the fundamental value having the dual
nature, connected both with current of external time, and with
properties of system.

\end{document}